\documentclass[aps,prl,twocolumn,superscriptaddress,groupedaddress]{revtex4}
\usepackage{subfig}
\usepackage{graphicx}  
\usepackage{dcolumn}   
\usepackage{bm}        
\usepackage{amssymb}   
\usepackage{slashed}
\usepackage{graphicx}				
\usepackage{amsmath}
\usepackage{mathtools}
\usepackage{tikz,pgf}
\usepackage{comment}
\usepackage{asymptote}
\usetikzlibrary{arrows,backgrounds}
\usetikzlibrary{fit,scopes,calc,matrix,positioning,decorations.pathmorphing}
\usepackage[all]{xy}
\usepackage{yfonts}
\newcommand{\bra}[1]{\ensuremath{\left\langle#1\right|}}
\newcommand{\ket}[1]{\ensuremath{\left|#1\right\rangle}}

\DeclareFontFamily{OMS}{oasy}{\skewchar\font48 }
\DeclareFontShape{OMS}{oasy}{m}{n}{%
         <-5.5> oasy5     <5.5-6.5> oasy6
      <6.5-7.5> oasy7     <7.5-8.5> oasy8
      <8.5-9.5> oasy9     <9.5->  oasy10
      }{}
\DeclareFontShape{OMS}{oasy}{b}{n}{%
       <-6> oabsy5
      <6-8> oabsy7
      <8->  oabsy10
      }{}
\DeclareSymbolFont{oasy}{OMS}{oasy}{m}{n}
\SetSymbolFont{oasy}{bold}{OMS}{oasy}{b}{n}

\DeclareMathSymbol{\smallleftarrow}     {\mathrel}{oasy}{"20}
\DeclareMathSymbol{\smallrightarrow}    {\mathrel}{oasy}{"21}
\DeclareMathSymbol{\smallleftrightarrow}{\mathrel}{oasy}{"24}

\begin{document}
\title{Qubit stabilisation via learning capable materials}
\author{Andrei T. Patrascu}
\begin{abstract}
I describe the engineered decoherence of a qubit state by means of an environment formed out of a neurally architected material. Such a material is a material that can adjust its inner properties in the same way a neural network is adjusting its weights, subject to a built-in cost function. Such a material is naturally found in biological structures (like a brain) but can in principle be engineered at a microscopic level. If such a material is used as an environment for a Nakajima-Zwanzig equation describing the controlled decoherence of a quantum state, we obtain a modified decoherence that allows for correlated states to exist longer or even to become robust. Such a neural material can also be architected to implement certain quantum gate operations on the encapsulated qubit. 
\end{abstract}
\maketitle
\section{Introduction}
Among the properties of various materials existing in nature, the learning capacity is often overlooked, and that with a good reason. Most materials we encounter in everyday life have only a tiny learning capacity, due to their dynamics being mostly non-neuronal. A crystallin structure would have a very hard time learning as it can hardly adapt its inner properties to some over-arching cost function. The same thing happens with most of the molecular materials and most of the other nano- or micro- structured materials. 
However, there exist cellular (or other engineered) materials that have an intrinsic neural structure [1-3], and for which learning comes easier. One is of course the neural tissue existing in all our brains. 
Such an environment may become an engineered bath for a quantum state that goes through a decoherence process through it. This process implies adding specific degrees of freedom to our quantum system to enable the determination in the environment of properties that were previously undefined, allowing for superposed states to be observed and ultimately perfectly and unambiguously determined. However, the key of engineered environments is to control precisely this addition of degrees of freedom in order to maintain some inner properties undetermined (hence in superposed states) for as long as possible [4-7]. It seems like an engineered neural network environment can naturally, by means of its dynamics, do exactly that. 
This would massively extend the level at which quantum superposed states can be achieved and maintained in otherwise unfavourable conditions (due to temperature, etc.) 
A physical implementation of a classical neural network is a physical system in which the dynamics of a classical neural network is performed by a medium that is to some extent decoherent. Such decoherence results in the apparently classical behaviour of the neurons (nodes) and of the signalling through the dendrites. The input and output signals of the neural network may be highly decoherent states from a quantum perspective. However, the inner dynamics of the neural network allows for enough superposition of states, and even of non-separability that entangled qubit states may be formed. 
It is important to observe that a "classical" physical neural network is not a purely classical neural network, for example as it would be implemented on a computer, but instead a decoherent state in which the complex phase, while existent, is hard to measure in interference observations. Therefore, I assume that no system is purely classical, but only asymptotically classical, and that inner mechanisms can amplify the coherence effects of apparently classical input states.

This must be mentioned in order to avoid a usual confusion between "emergence" of quantum mechanics from an underlying classical system, an idea I do not support here, and the transition towards a coherent state by means of the inner dynamics of a neural network, a point I am defending here, starting from a quantum, yet highly decoherent (therefore cvasiclassical) state. 

The general prescription for the emergence of asymptotically classical states is based on the phenomenon of decoherence, which by definition implies the partial cumulative entanglement of the system with its environment. Therefore, we notice that the quantum phase becomes undetectable because of a series of quantum phenomena that allow for information about the system to be measured within its environment. As such, there is no purely classical dynamics, only quantum dynamics that looks in certain limits very similar to what we would imagine to be classical. 

However, the type of interaction of our quantum (qubit) system with the environment is essential and can in fact be engineered such that certain properties of the qubit are not detectable by measuring the products of the interaction with the environment. Moreover, the interaction with the environment can also be engineered such that, after a series of interactions and after a certain amount of time passes, the outcome of a desired computation can be simply read by measuring the environment. 

The interactions with the environment can be designed such that they implement certain quantum operations on the qubits encoded in our system. This is done by the engineering of Liouville-Lindblad operators and their associated master equations. Normally, creating systems capable of doing that is not a simple experimental task. However, there exists a system that has the desired property of having a highly controlled type of coupling already implemented. That system is a neural network.

We consider the neural network at different stages in the computational process implementing different quantum computational requirements. Parts of the neural network will play the role of a superposed qubit state, while other parts will play the role of the environment. As the new "environment" will be that generated by a neural network, it can be adapted in order to implement required quantum operations while mitigating the uncontrolled decay of the inner states. This can be done by engineering the neural network without taking into account the decoherence processes that lie at the foundations of its operation. Therefore a cvasiclassical neural network can be engineered to perform quantum operations on a superposed, qubit like state maintained in the inner workings of a neural network or in an alternative neural network used as a qubit reservoir. 

In order to make some aspects of the discussion even clearer, we mention that a neural network simulation of a classical computer will be purely classical. In such a context we do not implement a quantum phase, so the system is not decoherent, but in fact purely classical. Such a simulated neural network would be of no use in the discussion of this article, as such an object would by definition have no complex phase and would be purely classical. In order to compute the effects I mention here, therefore, I will rely on a Hamilton-Jacobi type description of a neural network, combined with a proper description of its inner dynamics, without performing a simulation on a classical computer. This approach is capable of showing the theoretical underpinnings of this method and to prove that it would function as stated, but in order to experimentally verify it, one should use a physically implemented neural network. In order to simulate such a system one should either take a dynamical approach using Hamilton Jacobi type equations, or construct a simulation of a quantum, yet decoherent, neural network. The first approach is surely feasible, while the second one would obviously require significant computational capabilities, beyond my current possibilities. 


The construction of robust qubit states is particularly important for the creation of reliable and scalable quantum computers. Indeed, applying quantum operations on single or multiple qubit states is a practically difficult operation that must consider a series of experimentally intensive insulation and cooling procedures. This must be done in order to avoid producing a full determination of the inner state by the addition of new degrees of freedom to the system. Such an effect would lead to a change of the initial system that truly had some undetermined properties (say the projection of a qubit state) into one that fully determines the property of that state (say, the z projection of the qubit state) by means of environmental degrees of freedom. Keeping the state of a qubit in an undetermined state, allowing it to have all possible orientations as superpositions, for example, as described by its wavefunction implies some form of insulation from the environment. Such a robust representation could be achieved due to the inner workings of a physical neural network. Let us consider for example the input as a dynamical sequence of states provided to the neural network. Each neuron will fire according to its threshold (activation function), if the input is higher than a given threshold. The shape of the output of each neuron will also be determined by the activation function, leading to a signal emerging from the neuron and encoding some of the non-linearity of the activation function. The implementation of such functions in a quantum setting has been discussed in ref [10]. This signal will be distributed to a first, second, etc. layers of neurons, each adjusting its dynamics according to their activation function and to the signals they receive. They will receive multiple signals at each level, amounting to an inner superposition of the signals received by the previous layers. After each batch of data, a backwards optimisation process is started, and the inner parameters of the network are being optimised according to an extremum condition. While this creates a non-local, possibly non-separable state, the optimisation procedure can be adjusted such that the resulting outcome represents a robust combination of all initial states. If the input batch contains two electric signals (or two photon signals), the transfer of information into the neural network will provide us with an inner state of the network that can hold both input states (say spin up and spin down) in a superposition state, with a phase (or factor of the two ket vectors, for example) given by the various extremal configurations compatible with the cost function. Another way of saying would be that we transition from a single valued input, to a multiple valued result lying inside the neural network. A measurement of the outcome would amount to the extraction of an output of the neural network. However, this action is not required, and while the quantum computation is in process, not even desired. Keeping the network from providing an output state while keeping the dynamical state inside the network active amounts to keeping a robust qubit state inside the network. On this state we can then operate by means of modifications of the neural configuration that would maintain the initial non-determination. The final measurement would destroy the coherent state and would amount to providing an output of the network.
 
As a final comment to this introduction, it is interesting to note that even an apparently decoherent neural network may be capable of quantum computations, albeit not at the same level of efficiency as a quantum neural network itself, however, at temperature and environment conditions that would be deadly for a quantum neural network with all its inner components designed to remain quantum coherent. This makes it not impossible that highly dense, or even biological neural networks behave at least in some approximation as quantum computer emulators.

\section{Inner dynamics of neural networks and quantum coherence}
There is usually assumed that the same inner parameters of a neural network will face a sequence of inputs that will form a batch, leading to some optimisation that would adjust the parameters. This is usually insufficient when considering actual physical neural networks. In fact, in the best possible scenario, a neural network will have an inner dynamics and each input data will be presented with an inner structure of the neural network that is changed, both in terms of connectivity and in terms of weights, compared to what was presented to the first input. Such models are dynamical, and their dynamics allows to create interior superposed states corresponding to various input distributions. Such models present an inner dynamics that can adapt to various inputs. If one considers different inputs with slightly different properties and hence different computational needs, one performs the inference with different dynamical architectures adapted to each sample. This type of network can be used to dynamically re-route input data associated to different elements of the input distribution via multiple possible paths, preserving input state information across the network. This phenomenon leads to a multi-branch structure in which the path selection produces the phase structure we desire [11]. 

The dynamical neural network model is taken as the constructive example here. The neural network will need a component that will be able to alter the model behaviour at test or inference, and for each input given to the network. This will require an additional component traditionally called in the neural network literature a "Composer". Such a component will construct computational architectures in an adaptive way from sub-modules, given each input provided to the network. The focus will have to be on intermediate activations where a controller will be trained by reinforcement learning to find optimal intermediate activations. The controller policy itself can be altered at each test time (when input is provided). This allows us to keep track of several possible intermediate steps and to adapt the architecture and computation accordingly. Most importantly however, it encodes and maintains in the neural network different alternative representations associated to each of the inputs. In this way we obtain a construct in which we can have neural representations of qubits that are preserved robustly and encoded in a computationally efficient way. 
The robustness of the qubit implementation in a neural network can be seen by understanding decoherence as a result of the fact that the quantum system becomes entangled with its environment. That environment, if properly engineered, can be made to determine as little as possible (eventually nothing) about the superposed inner state, even when interactions exist, as the sole possible interactions will be non-revealing. 

It is important to make it clear here that the neural network is a physical one, in which some material, with an inner molecular or atomic structure has intrinsic quantum properties that may not emerge yet at a macroscopical level (it may not be a quantum material of any sort). The information provided at the input will be encoded by means of either electronic or laser pulses. While they will behave quasi-classical, their quantum component (their complex phase) will be existing. The representation of subsequent pulses in the neural network will lead to a superposition of states in the inner workings of the neural network, and hence inside the neural network we will have quantum states, not classical representations of quantum states. This is also mainly why a computer simulation of a purely classical neural network (as implemented nowadays in many a package) won't do the trick. 

The process of decoherence implies the fact that the environment is bringing enough suitable degrees of freedom in the system to allow the full determination of the property we desire to keep in a superposed (hence undetermined) state. With those degrees of freedom, the maximal knowledge about the system will fully determine its state and no superposition state will remain. 

We do however need superposed states in order to create suitable quantum computing states and to manipulate them later on. Therefore various approaches to preserving such superpositions have been implemented. On one side, the interaction with the environment doesn't necessarily destroy quantum coherence. There exist specific quantum states, depending on the type of interaction between the environment and the system, that lead to a conservation of the superposed states by the fact that the information available by measuring environment degrees of freedom after interactions do not reveal much or anything about the specific property of our system. 

Such "pointer" states can indeed be used to robustly encode quantum superposition states.  
Our quantum system of interest is a superposed state of a neural network, that is itself embedded into a larger neural network like material, making sure that the interactions of our system with the environment only happens through the neural material it is embedded into. 

In certain finite-dimensional systems it is often possible for the pointer states to form a preferred basis that we decide to call a pointer basis. The interaction with the environment mediated by an interaction hamiltonian $H_{int}$ imposes a dynamical filter on the state space, selecting those states that can be robustly prepared and observed even in the presence of the interactions with the environment. This mechanism induces an environment superselection rule. The system therefore will have to be defined by a system-focused Hamiltonian $H_{S}$, an environment focused Hamiltonian $H_{E}$ and an interaction focused Hamiltonian $H_{int}$. We wish to determine a set of system states $\{\ket{s_{i}}\}$ that remains unaffected by the interaction with the environment. In that sense, the states described above should remain unentangled with the environment under the evolution generated by $H_{int}$. The system-environment product state (assumed to be initially unentangled) will have to remain that way as time advances
\begin{equation}
e^{-i H_{int}t/\hbar}\ket{s_{i}}\ket{E_{0}}=\lambda\ket{s_{i}}e^{-i H_{int}t/\hbar}\ket{E_{0}}=\ket{s_{i}}\ket{E_{i}(t)}
\end{equation}
In the limit in which the interaction part of the Hamiltonian is also the dominant part $H\sim H_{int}$, also known as the quantum measurement limit, those pointer states are obtained by diagonalising $H_{int}$ in the subspace of the system. 
From an operatorial point of view, we may as well define a pointer observable
\begin{equation}
\mathcal{O}_{S}=\sum_{i}o_{i}\ket{s_{i}}\bra{s_{i}}
\end{equation}
in the form of a linear combination of pointer state projectors $\Pi_{i}=\ket{s_{i}}\bra{s_{i}}$. Because each pointer state is an eigenstate of the interaction hamiltonian it follows that 
\begin{equation}
[\mathcal{O}_{S}, H_{int}]=0
\end{equation}
We have already a good measure for what a robust state would look like, at least from the perspective of a simply implemented quantum state. What has to be shown is that if the input of a physically implemented neural network is a quantum state, due to the dynamics of the neural network, it will turn into a pointer state of the neural network hamiltonian, given the types of interactions allowed between the neural network and the environment. There will be a topological component to the state encoded via a neural network, that results on how the signals are being transmitted. However, the way in which decoherence is mitigated by a neural network induced superposition will play an interesting role in understanding the interaction between a neural system and its environment. Basically, if interactions with the environment of the neural network are considered to occur through the neural network boundary, the encoding of the quantum states within the neural network will have the properties required for the states to be robust with respect to $H_{int}$. 

Therefore, neural networks will allow for a robust encoding of quantum information. 
With this, we may construct a neural network that, while being decoherent at the level of the signal transmission, has a dynamics (the usual dynamics of a neural network) leading to an environment that minimises decoherence and moreover, allows the reading of the outcome of the quantum computation via the neural network seen as an environment. In this sense, the neural network constitutes a type of advanced environment engineering, in which the quantum operations are performed on the input by the inner workings of the network, and the result is read in the environment that is made out of the neural network. In this way I propose an engineering of the environment that, as opposed to ref [1] is this time dynamical, obeying the dynamical laws of a neural network. Reading out the result of the computation is done within the neural network time register with a classical probability that depends inversely proportional on the total time steps of the register. 

This type of neuronal aided quantum computing, in which the environment engineering becomes dynamical and encoded via a neural network is fundamentally new, but can indeed be practically realised due to the recent advancements in creating neural materials, that is, materials that have their microscopic constituents engineered such that they can perform the functions of a neural network also known as learning materials [2]. 

From a mathematical standpoint, decoherence is encoded by means of a master equation which is given by a density matrix dependent master operator that obeys a Liouvillian type equation in Lindblad form 
\begin{equation}
\mathcal{L}(\rho)=\sum_{k}L_{k}\rho L_{k}^{\dagger}-\frac{1}{2}\{L_{k}^{\dagger}L_{k},\rho\}_{+}
\end{equation}
We need to find such a master equation where the operators $L_{k}$ act locally, and the equation has a steady state $\rho_{0}$ that is unique and attainable in a polynomial time. 

The neural network receives a series of electric or light pulses as inputs, but the neural network is itself seen as a decohered system from the point of view of its inner connections, albeit not from the point of view of the dynamics of the signals. Can such a neural environment (that is for all practical purposes classical) be used to not only maintain the robustness of the quantum input information, but also to amplify its quantum nature and to robustly process it and retrieve computational results? I believe the answer is yes. 

\section{quantum computing}
In the previous section I showed that a dynamical neural network can adaptively maintain different alternative paths providing an actual superposition of states, similar to that of a qubit. However the system still looks at least semi-classical as it remains highly decoherent and while the system does hold different paths, the properties of each path can be ultimately perfectly determined, hence the system will not be able to provide coherent operations. This however would be the case if the input signal was strictly classical. However, in a physical neural network, the input is never truly classical, and the quantum phase will always exist, even if it wouldn't initially have a clear visible quantum property. We have to consider therefore each path created by the composer as taking a decoherent input, and via a series of optimisation tools provided by the combination of the dynamics of the composer and the controller, we will obtain an amplification of the coherence of the input on each trajectory. The result would be that of a quantum coherent state in which a quantum coherent superposition of states will emerge. This is not a classical to quantum transition, but rather an amplification of the quantum nature by a series of optimisations provided by the neural network and its components. 
Following Feynman's construction of a quantum simulator, we will represent the time as an auxiliary register given by the quantum states $\{\ket{t}\}_{t=0}^{T}$ and the Lindblad operators will be 
\begin{equation}
L_{i}=\ket{0}_{i}\bra{1}\times \ket{0}_{t}\bra{0}
\end{equation}
and in terms of time development 
\begin{equation}
L_{t}=U_{t}\times \ket{t+1}\bra{t}+U_{t}^{\dagger}\times \ket{t}\bra{t+1}
\end{equation}
The two indices count $i=1...N$ and $t=0...T$. 
The neural network will be encoded in the form of the Liouville operator which basically will contain the coupling terms of various layers of neurons as linear combinations of terms and coupling parameters associated with the interactions between neurons in a layer. The neural network dynamics involving both its geometry and topology will be manifest in the dynamical reorganisation of the Liouville operator as time advances. That will lead to potentially several steady states corresponding to different geometries and inner network configurations. 
\begin{equation}
\rho_{0k}=\frac{1}{T+1}\sum_{t}\ket{\psi_{t}}\bra{\psi_{t}}\times \ket{t}\bra{t}
\end{equation}
where $\mathcal{L}(\rho_{0k})=0$. Each such state is unique for a given configuration, but can be altered by modifying the neural network. Indeed, as shown above, the composer will construct different neural architectures adaptively, given each input provided to the network, allowing for intermediate activations by a controller that will find optimal intermediate activations. Therefore we will control several superposed states at the same time, as well as the possibility of coupling the inner neural states to different layers of the neural network in controlled ways. 
In general we can regard the Lindblad equation as
\begin{equation}
\frac{d\ket{\rho}}{dt}=\mathcal{L}\ket{\rho}
\end{equation}
which we can expand in the form of 
\begin{equation}
\frac{d\rho}{dt}=-i[H,\rho]+\sum_{k}(L_{k}\rho L_{k}^{\dagger}-\frac{1}{2}\{L_{k}^{\dagger}L_{k},\rho\})
\end{equation}
where the summation is over all possible channels of environmental interaction. Those channels will form the dynamics of the neural network encoding the inner dynamics, including superpositions, interactions and propagation inside the neural network. 

\begin{equation}
\mathcal{L}=-i \mathbb{I}\otimes H + i H \otimes \mathbb{I}+\sum_{k}L_{k}^{*}\otimes L_{k}-\frac{1}{2}\mathbb{I}\otimes (L_{k}^{\dagger}L_{k})-\frac{1}{2}L_{k}^{T}L_{k}^{*}\otimes \mathbb{I}
\end{equation}
The Lindbladian above can be used to generate the time propagation as in 
\begin{equation}
\ket{\rho(t)}=e^{\mathcal{L}t}\ket{\rho(0)}
\end{equation}
The evolution of the off diagonal elements of the density matrix will encode the quantum coherence and its vanishing into the environment. In order to encode this in the case of a neural network environment (or the process of decoherence in a learning material) we encode the learning dynamics of the neural network by means of a series of activation functions acting on the impulses received by each layer. Having a dynamical neural network that preserves several paths for the states at the same time, the superposition will be considered among those states. The activation dynamics is governed in this example by a hyperbolic tangent function which will be applied as the dynamical part of the construction of the Lindblad operator. 
 The interaction hamiltonian will be of the form encoding nearest neighbours interaction, as a connection between signals from different layers. 
The Lindbladian can be written as an operator that implements the transformations produced by a neural network given a decoherent input. 
As opposed to the usual applications of the Lindblad operators, a neural network would consist in a non-Markovian model which is characterised by extensive memory effects. This amounts to a system with strong couplings to the environment, correlation and entanglement in the initial state, and even the equivalent of interactions with the environment at low temperatures. 

The general approach to such systems is the projection operator technique in which we use projection superoperators $\mathcal{P}$ acting on states of the total system including the environment. This superoperator represents the elimination of certain degrees of freedom form the complete description of the states of the total system. If $\rho$ describes the full composite system, the projection $\mathcal{P}\rho$ represents an approximation of $\rho$ leading to a simplified effective description of $\rho$ which leads to a simplified effective description of the dynamics given a reduced set of variables. This projection is called the relevant part of the density matrix. It defines a closed set of dynamical equations for the relevant part of the density matrix. The resulting equation is a Nakajima-Zwanzig equation, an integro-differential equation with a kernel containing retarded memory effects. A simplification is a time-convolutionless master equation which is a time-local differential equation of first order involving a time-dependent generator. From one of these equations we then arrive at effective master equations by various tools, mainly expanding in a perturbative series with respect to the system-environment coupling. 
In non-Markovian processes, one usually searches for the form of the projected density matrix as a separable product of the system matrix and the fixed initial environment
\begin{equation}
\mathcal{P}\rho = \rho_{S}\otimes \rho_{0}
\end{equation}
with $\rho_{S}=tr_{E}\rho$ the reduced matrix of the open system. 
This system does take the correlations with the environment only in a perturbative way into account. This is sufficient in many situations, but it cannot be enough if the environment is a learning material. This is precisely because a learning material has memory as a defining property, allowing a very specific type of adaptation to previous input, and hence playing the role of a strongly correlated channel. Therefore we need a superoperator that acts like
\begin{equation}
\mathcal{P}\rho=Tr_{E}(A_{i}\rho)\otimes B_{i}
\end{equation}
with the same definitions as above. The operators $A_{i}$ and $B_{i}$ satisfy 
\begin{equation}
tr_{E}\{B_{i}A_{j}\}
\end{equation}
and 
\begin{equation}
\sum_{i}(tr_{E}B_{i})A_{i}=I_{E}
\end{equation}
as well as 
\begin{equation}
\sum_{i}A_{i}^{T}\otimes B_{i}\geq 0
\end{equation}
In general there are many different sorts of operators $A_{i}$ and $B_{i}$ that represent a given superoperator $\mathcal{P}$. If we have a particular set of such operators then we can represent the same projection by transforming 
\begin{equation}
\begin{array}{c}
A_{i}'=\sum_{j}u_{ij}A_{j}\\
B_{i}'=\sum_{j}v_{ij}B_{j}\\
\end{array}
\end{equation}
where the matrices $u$ and $v$ are real non-singular matrices satisfying $u^{T}v=I$. 
Given a projection superoperator we may define the relevant states as being those in the range of $\mathcal{P}$ for which the relation 
\begin{equation}
\mathcal{P}\rho_{rel}=\rho_{rel}
\end{equation}
holds. 
These states are of the form 
\begin{equation}
\rho_{rel}=\sum_{i}\rho\otimes B_{i}
\end{equation}
with $\rho_{i}$ being positive matrices such that $\sum_{i}tr_{S}\rho_{i}=1$. The manifold of the relevant states is determined by the operator $B_{i}$. In terms of observables we may write that a hermitian operator $\mathcal{O}_{rel}$ on the total state space is relevant if 
\begin{equation}
tr\{\mathcal{O}_{rel}(\mathcal{P}\rho)\}=tr\{\mathcal{O}_{rel}\rho\}
\end{equation}
for all $\rho$. The expectation value of a relevant observable in any state of the composite system is left unchanged by the superoperator. Given the adjoint projection operator this means
\begin{equation}
\mathcal{P}^{\dagger}\mathcal{O}_{rel}=\mathcal{O}_{rel}
\end{equation}
hence the relevant observables are invariant under adjoint projection, but given 
\begin{equation}
\mathcal{P}\rho=\sum_{i} tr_{E}\{A_{i}\rho\}\otimes B_{i}
\end{equation}
we obtain 
\begin{equation}
\mathcal{P}^{\dagger}\mathcal{O}=\sum_{i}tr_{E}\{B_{i}\mathcal{O}\}\otimes A_{i}
\end{equation}
and hence the relevant observables must be 
\begin{equation}
\mathcal{O}_{rel}=\sum_{i}\mathcal{O}_{S}^{i}\otimes A_{i}
\end{equation}
which means that the structure of the relevant observables is determined by $A_{i}$. 
If we denote the unitary time evolution operator by $U(t)$ then we can write 
\begin{equation}
\rho(t)=U(t)\rho(0)U^{\dagger}(t)
\end{equation}
as describing the dynamics of the state. But this we may restrict by means of the projector superoperator to relevant sates and the associated operators. Due to the operator nature of the density and it being characterised by the $A_{i}$ operators we will write the dynamical variables as 
\begin{equation}
\rho_{i}(t)tr_{E}\{A_{i}\rho(t)\}
\end{equation}
and by using the positivity of those variables as well as the normalisation $tr_{E}B_{i}=1$ we obtain the reduced density matrix of the system as 
\begin{equation}
\rho_{S}(t)=\sum_{i}\rho_{i}(t)
\end{equation}
The state of the reduced system is determined by a certain set of unnormalised density matrices $\rho_{i}(t)$. If the initial state belongs to the manifold of relevant states 
\begin{equation}
\rho(0)=\mathcal{P}\rho(0)=\sum_{i}\rho_{i}\otimes B_{i}
\end{equation}
then the Nakajima-Zwanzig equaton has no inhomogeneity although the initial state is one that has state-environment correlations. The dynamics will then lead to 
\begin{equation}
\rho_{i}(t)=\sum_{j} tr_{E}\{A_{i}U(t)\rho_{j}(0)\otimes B_{j}U^{\dagger}(t)\}
\end{equation}
We bring those dynamical variables into a vector 
\begin{equation}
\eta=(\rho_{1}, \rho_{2}, ..., \rho_{n})
\end{equation}
and then define the dynamical transformation 
\begin{equation}
V_{t}: \eta(0)\rightarrow \eta(t)
\end{equation}
This is a one-parameter family of maps reproducing the dynamics of the system, where $V_{0}$ is the identity map. This map is not an operation on the space of reduced systems, but a map on the space of vectors $\eta$. The transition $\eta(0)\rightarrow \rho_{S}(0)=\sum_{i}\rho_{i}(0)$ comes with a loss of information on the initial correlations and therefore if we know $\rho_{S}(0)$ we cannot derive the dynamical behaviour in general. There is no prescription that assigns to each $\rho_{S}(0)$ a unique $\rho_{S}(t)$.
The description of you open system will now be described by means of a closed dynamic equation governing the relevant projected total density matrix 
\begin{equation}
\frac{d}{dt}\mathcal{P}\rho(t)=\mathcal{K}^{t}(\mathcal{P}\rho(t))
\end{equation}
where $\mathcal{K}^{t}$ is a linear generator. 
Given 
\begin{equation}
\mathcal{P}\rho=\sum_{i} tr_{E}\{A_{i}\rho\}\otimes B_{i}
\end{equation}
we obtain a system of equations
\begin{equation}
\frac{d}{dt}\rho_{i}=\mathcal{K}_{i}^{t}(\rho_{1},...,\rho_{n})
\end{equation}
for $i=1,2,...,n$. The generator is explicitly time dependent and each one depends linearly on the input dynamical arguments $\rho_{i}$. 

The process described up to now relies on the construction of the Nakajima-Zwanwig equation where a projection operator $\mathcal{P}$ was used on the state space of the total system, having the property of decomposing the system into a relevant part of the total density matrix described by $\mathcal{P}\rho$ which obeys some equations of motion and corresponds to the slowly fluctuating memory kernel part, and a so called irrelevant or rapidly fluctuating part. The memory function of the Nakajima-Zwanwig equation is determined by the Mori-Zwanwig equation and combined with the fast fluctuating part provides a full description of the system.

The situation becomes interesting if we embed our quantum state (say a qubit) into an environment that is basically an engineered bath that has all the properties of a classical neural network. This behaviour gives its bath the properties of microscopically architected neural networks. Such a material is one that has its microscopic components designed to perform learning and optimisation tasks at the level of its fundamental constituents. Such materials are capable of changing their inner properties in the way in which artificial neural network tune their weights according to in-built cost functions. While designing such materials may be complicated, we know of various biological materials that behave in quite a similar fashion. In any case, if the environment of our quantum state is designed to play the role of such a material, hence we have an engineered bath that obeys the dynamics of a neural network, being therefore a highly non-markovian construct, we have to write the master equations with specific components for the generator terms. 

The dynamical equation is expected to preserve the positivity of all components $\rho_{i}$. We introduce an auxiliary Hilbert space and an orthonormal basis $\{\ket{i}\}$ on this space. We expand our vector $\eta$ to be defined on this extended space as 
\begin{equation}
\eta=\sum_{i}\rho_{i}\otimes \ket{i}\bra{i}
\end{equation}
The auxiliary space is the construction allowing us to analyse the additional degrees of freedom describing the statistical correlations induced by the projection superoperator $\mathcal{P}$. The extended density matrix is a block diagonal matrix with the blocks $\rho_{i}$ along the main diagonal. The reduced matrix $\rho_{S}$ is obtained by partial trace over the auxiliary space. The dynamical transformation preserves the block diagonal structure. For this to happen we need a Lindblad generator on the extended space with the property that 
\begin{equation}
\mathcal{L}(\sum_{i}\rho_{i}\otimes \ket{i}\bra{i})=\sum_{i}\mathcal{K_{i}}(\rho_{1},...,\rho_{n})\otimes \ket{i}\bra{i}
\end{equation}
The solution of the dynamical equation is then 
\begin{equation}
\sum_{i}\rho_{i}(t)\otimes \ket{i}\bra{i}=e^{\mathcal{L}t}(\sum_{i}\rho_{i}(0)\otimes \ket{i}\bra{i})
\end{equation}
This induces a special form for the generators $\mathcal{K}_{i}$ as time independent 
\begin{equation}
\mathcal{K}_{i}(\rho_{1},\rho_{2},...,\rho_{n})=-i[H^{i},\rho_{i}]+\sum_{j\lambda}(S_{\lambda}^{ij}\rho_{j}S_{\lambda}^{ij\dagger}-\frac{1}{2}\{S_{\lambda}^{ij\dagger}S_{\lambda}^{ij},\eta \})
\end{equation}
where
\begin{equation}
\begin{array}{c}
H=\sum_{i}H^{i}\otimes \ket{i}\bra{i}\\
\\
S_{\lambda}^{ij}=R_{\lambda}^{ij}\otimes \ket{i}\bra{j}\\
\end{array}
\end{equation}
The system in contact with its environment will then be described by a hamiltonian 
\begin{equation}
H=H_{S}+H_{E}+V
\end{equation}
where we consider the Hamiltonian of our system to be a standard two-state system defined by means of lowering and rising operators 
\begin{equation}
H_{S}=Q\cdot \sigma_{+}\sigma_{-}
\end{equation}
The environment must be described in terms of the dynamics of a neural network.
Such a dynamics allows only specific channels for information output, in particular, as a superposition of the initial input states by means of linear combinations, with weights dynamically adapted to the cost function of the neural network. Those linear combinations then are being transferred into an activation function, which here will be considered to be the Hyperbolic tangent. Using the Hamilton Jacobi approach to the learning dynamics of neural networks, as described in ref. [8] we divide the learning process into epochs constituting a certain number of batches leading to an update of the weights. 
The observables associated to the neural networks become functions of time and of the weights 
\begin{equation}
J(t)=J(t,y(t),W(t))
\end{equation}
and the dynamics of the observables has the geometrical interpretation of a single surface described by the equation 

\begin{equation}
D(t, y, W, \frac{\partial J}{\partial t}, \frac{\partial J}{\partial y}, \frac{\partial J}{\partial W})=0
\end{equation}
If we associate conjugate variables to the weights $W$ which are themselves becoming dynamical variables, call those $M=\frac{\partial J}{\partial W_{ij}}$ and calling $\Delta_{i}=\frac{\partial J}{\partial y_{i}}$ we have the Hamilton Jacobi equation 
\begin{equation}
\frac{\partial J}{\partial t}+H(t,y,\Delta, W, M)=0
\end{equation}
with $J$ playing the role of the action functional and $y(t)$ representing the neuron output described by the activation dynamics in the following form 
\begin{equation}
\frac{d}{dt}y(t)=F(input(t), y(t), W(t))
\end{equation}

Given a past epoch of length T, we have a recursive decomposition of the weight functions as
\begin{equation}
W_{ij}(nT+\tau)=W_{ij}((n-1)T + \tau)+\Delta W(nT+\tau)
\end{equation}
with $t=n\cdot T+ \tau$ where $\Delta W$ is the change in the weights experienced during an epoch. 
The Hamiltonian associated to this decomposition has the form 
\begin{equation}
H=\sum_{k}\Delta_{k}\cdot F_{k}(t,y,S_{T}W)+\frac{1}{2\omega}\sum_{k,l}\sum_{v=0}^{n+1}(S_{vT}M)^{2}_{kl}+E(t,y,W)
\end{equation}
where 
\begin{equation}
F_{k}=\frac{1}{\lambda}[-y_{k}+f_{k}(\sum_{j=-1}^{N}S_{T}W_{kj}\cdot y_{j})]
\end{equation}
and
\begin{equation}
(S_{vT}X)_{kl}(t)=X_{kl}(t-vT)
\end{equation}
where $X=(W,M)$. 
The function $E(t, y, W)$ can be seen as an associated error function, or a generalised cost function for the neural material. The function $f$ reflects the topology of the network and encodes the time constant of a neuron by means of the constant $\frac{1}{\lambda}$. The two timescales here, will be associated with the weights dynamics timescale determined by the activation function and the learning process across various weights batches leading to updating the weights to optimise the cost function. 
The input of the environment network, from the inner quantum state is given in a form that combines the coherent state of the network representation of the qubit with the engineered form of the environment. 
The cost function represents the potential directing the evolution of the system as it enters in contact with the environment, and hence its minimisation by means of standard neural dynamics leads to a robust coherent inner qubit state. In general this will represent a sequential coupling of the two qubit states and can be represented here as 
\begin{equation}
V=\frac{1}{\lambda}\sum_{n_{1},n_{2}}c(n_{1},n_{2})\sigma_{+}\ket{n_{1}}\bra{n_{2}}+h.c.
\end{equation}
where the indexes label the two possible regions for the quantum state. 

The neural network can also be described in terms of a Kernel representation as every model learned by gradient descent is approximately a Kernel Machine [9] therefore the neural network decoherence environment is amenable to a kernel description a la Nakajima Zwanzig. In the most general context we apply the projection operators $\mathcal{P}$ and $\mathcal{Q}$ that extract the relevant and irrelevant parts of the density matrix and hence produce 
\begin{widetext}
\begin{equation}
\begin{array}{c}
\frac{\partial}{\partial t}\mathcal{P}\rho(t)=\mathcal{P}\frac{\partial}{\partial t}\rho(t)=\alpha \mathcal{P}L(t)\rho(t)\\
\\
\frac{\partial}{\partial t}\mathcal{Q}\rho(t)=\mathcal{Q}\frac{\partial}{\partial t}\rho(t)=\alpha \mathcal{Q}L(t)\rho(t)\\
\end{array}
\end{equation}
\end{widetext}
where $L$ is a Liouville operator. We can rewrite the identity operation as $\mathcal{I}=\mathcal{P}+\mathcal{Q}$ and obtain 

\begin{widetext}
\begin{equation}
\begin{array}{c}
\frac{\partial}{\partial t}\mathcal{P}\rho(t)=\alpha\mathcal{P}L(t)\cdot \mathcal{I}\cdot \rho(t)=\alpha \mathcal{P}L(t)(\mathcal{P}+\mathcal{Q})\rho(t) =\alpha \mathcal{P} L(t)\mathcal{P}\rho(t)+\alpha \mathcal{P}L(t)\mathcal{Q}\rho(t)\\
\\
\frac{\partial}{\partial t}\mathcal{Q}\rho(t)=\alpha\mathcal{Q}L(t)\cdot \mathcal{I}\cdot \rho(t)=\alpha \mathcal{Q}L(t)(\mathcal{P}+\mathcal{Q})\rho(t) =\alpha \mathcal{Q} L(t)\mathcal{P}\rho(t)+\alpha \mathcal{Q}L(t)\mathcal{Q}\rho(t)\\
\end{array}
\end{equation}
\end{widetext}
We search a formal solution corresponding to a given initial density matrix 
\begin{widetext}
\begin{equation}
\mathcal{Q}\rho(t)=K(t,t_{0})\mathcal{Q}\rho(t_{0})+\alpha\int_{t_{0}}^{t}ds K(t,s)\mathcal{Q}L(s)\mathcal{P}\rho(s)
\end{equation}
\end{widetext}
where the propagator is 
\begin{equation}
K(t,s)=\mathcal{T}exp[\alpha\int_{s}^{t}ds'\mathcal{Q}L(s')]
\end{equation}
ordered chronologically. 
The propagator satisfies the differential equation given by 
\begin{equation}
\frac{\partial}{\partial t}K(t,s)=\alpha \mathcal{Q} L(t) K(t,s)
\end{equation}
with $K(s,s)=\mathcal{I}$. 
With this we obtain 
\begin{widetext}
\begin{equation}
\begin{array}{c}
\frac{\partial}{\partial t}\mathcal{P}\rho(t)=\alpha \mathcal{P}L(t)\mathcal{P}\rho(t)+\alpha \mathcal{P}L(t)[K(t,t_{0})\mathcal{Q}\rho(t_{0})+\alpha\int_{t_{0}}^{t} ds K(t,s)\mathcal{Q}L(s)\mathcal{P}\rho(s)]=\\
\\
\alpha \mathcal{P}L(t)\mathcal{P}\rho(t)+\alpha \mathcal{P}L(t)K(t,t_{0})\mathcal{Q}\rho(t_{0})+\alpha^{2}\int_{t_{0}}^{t} ds K(t,s)\mathcal{Q}L(s)\mathcal{P}\rho(s)\\
\end{array}
\end{equation}
\end{widetext}
This projection does not represent a simple product state, and hence it projects onto correlated states, leaving us to construct our model assuming strong correlations between the system $S$ and its environment. 
With this in mind, and a Kernel construction as the one provided in ref [x] via the Hamilton Jacobi equation leads a proper way in which we can write the environment together with the correlation between our system and the environment. 

I will first analyse the evolution of the density matrix in the context of a normal Lindblad equation with no additional neural structure. 
To model however the proper evolution of interacting (coupled) signals the evolution equation will involve coupling between non-diagonal (coherence) matrix elements of one region and the diagonal (decoherent) elements of the other in the evolution of the density matrix.
\begin{widetext}
\begin{equation}
\frac{d\rho_{ij}}{dt}=\frac{1}{2}(\rho_{ij}(t)+\rho_{i j-1}(t)+\rho_{i-1 j}(t)\rho_{ij}(t)+\rho_{i-1 j}(t)\rho_{i j-1}(t)-\rho_{i-1 j}(t)\rho_{i j-1}(t)\rho_{ij}(t))
\end{equation}
\end{widetext}
This evolution will be compared with the one resulting from the successive application of the hyperbolic tangent activation function to the superposition of the next layer into the equation 
\begin{widetext}
\begin{equation}
\frac{d\rho_{ij}}{dt}=\frac{1}{2}(\rho_{i j-1}(t)+\rho_{ij}(t)+tanh(\rho_{ij}(t)+\rho_{i j-1}(t)+\rho_{i-1 j}(t)\rho_{ij}(t)+\rho_{i-1 j}(t)\rho_{i j-1}(t)-\rho_{i-1 j}(t)\rho_{i j-1}(t)\rho_{ij}(t)))
\end{equation}
\end{widetext}
The result of the first evolution is shown in figure 1. We can observe how the first equation produces off diagonal elements that quickly vanish as the system evolves. When the activation function and additional neural layers are being implemented, as seen in figure 2, the off-diagonal terms do not fall off, and we detect a phase structure relevant for maintaining coherence. Further on, we investigate also the entanglement of the various states introduced in the different paths analysed in the neural network. For that we use an entanglement measure. The evolution is shown in figure 3. There In the usual Lindblad case, we notice a quick decay of the entanglement of the encoded quantum states. However, if a neural environment is present, the quantum coherence is maintained, and our entangled state is preserved within the two quantum states. The neural network plays therefore the role of a dynamical engineered environment that not only increases the robustness of the quantum states, it also enhances coherence and is capable of implementing quantum gate operations.
\begin{figure}
  \includegraphics[width=\linewidth]{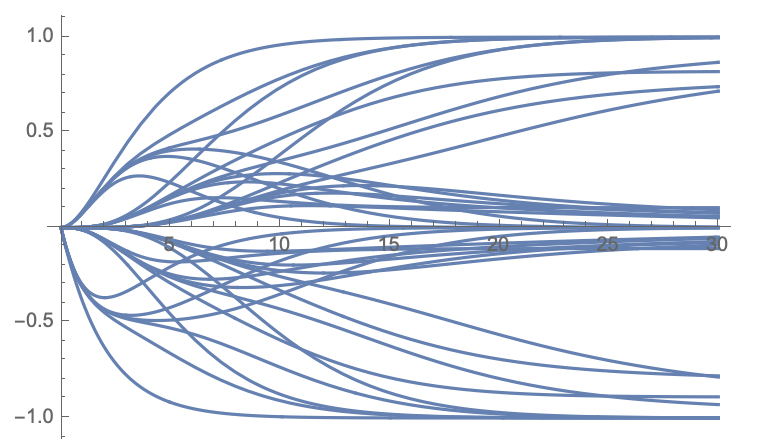}
  \caption{evolution of the diagonal and off diagonal density matrix elements in a non-engineered environment which is not a learning material}
  \label{graph1}
\end{figure}

\begin{figure}
  \includegraphics[width=\linewidth]{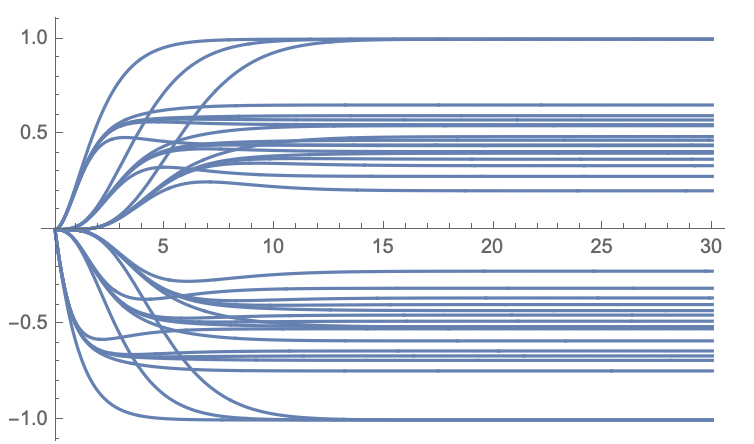}
  \caption{evolution of the diagonal and off diagonal density matrix elements in a neural learning material, a material capable of implementing a learning dynamics according to a cost function represented in a Hamiltonian form}
  \label{graph1}
\end{figure}
\section{conclusion}
I showed that if a superposed quantum state is in contact with an engineered environment that manifests the dynamics of a neural material, the coupling with that environment does not reduce the coherence of the superposed quantum state. This would facilitate not only the creation of a resource for quantum computation, but would also allow to operate on the quantum state in order to perform gate-type operations. There are obvious difficulties: the neural network surrounding the quantum state would have to be artificially structured at a level at which the interaction with the superposed inner state would follow the specific channels of a neural network, and it would have to be following only those channels. Also the construction of neural materials having precisely the properties of decoherent (classical) neural networks is not experimentally simple at this point. However, if this can be achieved, a physical neural network can play the role of an engineered environment that could turn a superposed quantum state into a robust state. 
\section{Statements and Declarations}
I declare there is no conflict of interest in this work. There is no relevant new data that has been generated during this work.

\end{document}